\documentclass[11pt]{article}
\usepackage{epsfig}

\begin{document} 

\title{Critical exponents in stochastic sandpile models} 

\author{Alessandro Chessa$^1$, Alessandro Vespignani$^2$\\ 
and Stefano Zapperi$^3$\\ \ \\
$^1$ Dipartimento di Fisica and Unit\'a INFM, Universit\'a di Cagliari,\\ 
Via Ospedale 72, 09124 Cagliari, Italy\\
$^2$ The Abdus Salam International Centre for Theoretical Physics (ICTP),\\ 
P.O. Box 586, 34100 Trieste, Italy  \\
$^3$ PMMH-ESPCI, 10 Rue Vauquelin, 75234 Paris CEDEX 05, France }

\date{}
\maketitle

\begin{abstract}
We present large scale simulations of a stochastic sandpile model in 
two dimensions. We use moments analysis to evaluate critical exponents  
and finite size scaling method to consistently test the obtained results. 
The general picture resulting from our analysis allows us to characterize 
the large scale behavior of the present model with great accuracy.  
\end{abstract}

Sandpile automata \cite{btw} are prototypical models 
to describe avalanche transport processes. 
All these models show a stationary state that after a suitable
tuning of the driving fields\cite{vz} displays a singular response 
function characterized by power law distributed events. These distributions
are typically bounded by upper cut-offs related to the system size.
In analogy with critical phenomena, 
is possible to define a complete set of scaling exponents 
describing the large scale behavior of these  models. 

Despite the large conceptual impact and the  huge effort devoted 
to the study of  sandpile automata in the last ten years, many 
basic issues, such as the precise values of the critical exponents,
the identification of universality classes and of the
upper critical dimension, still lay unresolved.
Theoretically, many approaches \cite{vzp,diaz,ron}  point out that 
different sandpile models, such as the Bak, Tang and 
Wiesenfeld (BTW) \cite{btw} and the Manna
\cite{manna} models, all belong to the same universality class.
Theoretical estimates for  critical exponents have been provided 
(especially in Euclidean dimension $d=2$) by means of different methods
\cite{vzp,diaz,prie}, and some exact results \cite{exact} can be derived 
from the Abelian structure of the BTW model. 
Numerical results are difficult to interpret, since different methods 
of analysis typically yield different results
\cite{manna,grasma,ben,att,cmvz,lubeck}. This is probably due to intrinsic 
scaling anomalies and finite size effects present in sandpile models. 

Here, we present very large scale numerical simulations of the 
Manna model\cite{manna}, that is the standard example of 
a sandpile automaton with stochastic toppling rule. We show that 
Manna model can be coherently described within a finite size scaling 
(FSS) framework. Critical exponents are evaluated with great accuracy 
and the results are confirmed by data collapse analysis. 

We consider a two-dimensional square lattice of linear size $L$
and associate to each site an integer variable $z_i$ (energy).
At each time step an energy grain is added on a randomly chosen site
($z_i\to z_i+1$). When one of the sites reaches or exceeds the 
local  threshold $z_c=2$ a ``toppling'' occurs:: $z_i=z_i-2$ and $z_j=z_j+1$,
where $j$ represents two randomly chosen nearest neighbor sites of site $i$.
A toppling
can induce nearest-neighbor sites to topple on
their turn and so on, until all the lattice 
sites are below the critical threshold.
This process is called an avalanche.  
Grains are added only when all the sites are below the threshold,
which corresponds to a fine tuning of the external driving field \cite{vz}. 
In addition, the model is conservative 
and energy is dissipated only at boundary
sites \cite{btw,manna}.

Avalanches in sandpile models are usually characterized by three
variables: the number of topplings $s$, the area $a$ affected
by the avalanche, and the avalanche duration $t$. 
The probability distribution of each of
these variables can be described as a power law with a cutoff
\begin{equation}
P(x)=x^{-\tau_x} {\cal G}(x/x_c),
\label{pl}
\end{equation}
where $x=s,a,t$. When
the system size $L$ goes to infinity 
the cutoff $x_c$ diverges as $x_c\sim L^{\beta_x}$. 
Under the finite size scaling (FSS) assumption of Eq.~(\ref{pl}), the 
set of exponents $\{\tau_x,\beta_x\}$ defines the universality
class of the model. 
In order to test the above FSS picture and to find an accurate estimate
of the various critical exponents,
we perform numerical simulations of 
two-dimensional  Manna model with open boundary conditions 
and conservative dynamics. The lattice size ranges 
from $L=128$ to $L=2048$, and statistical distributions are 
obtained averaging over $10^7$ nonzero avalanches. 
\begin{figure}[t]
  \centerline{\epsfig{file=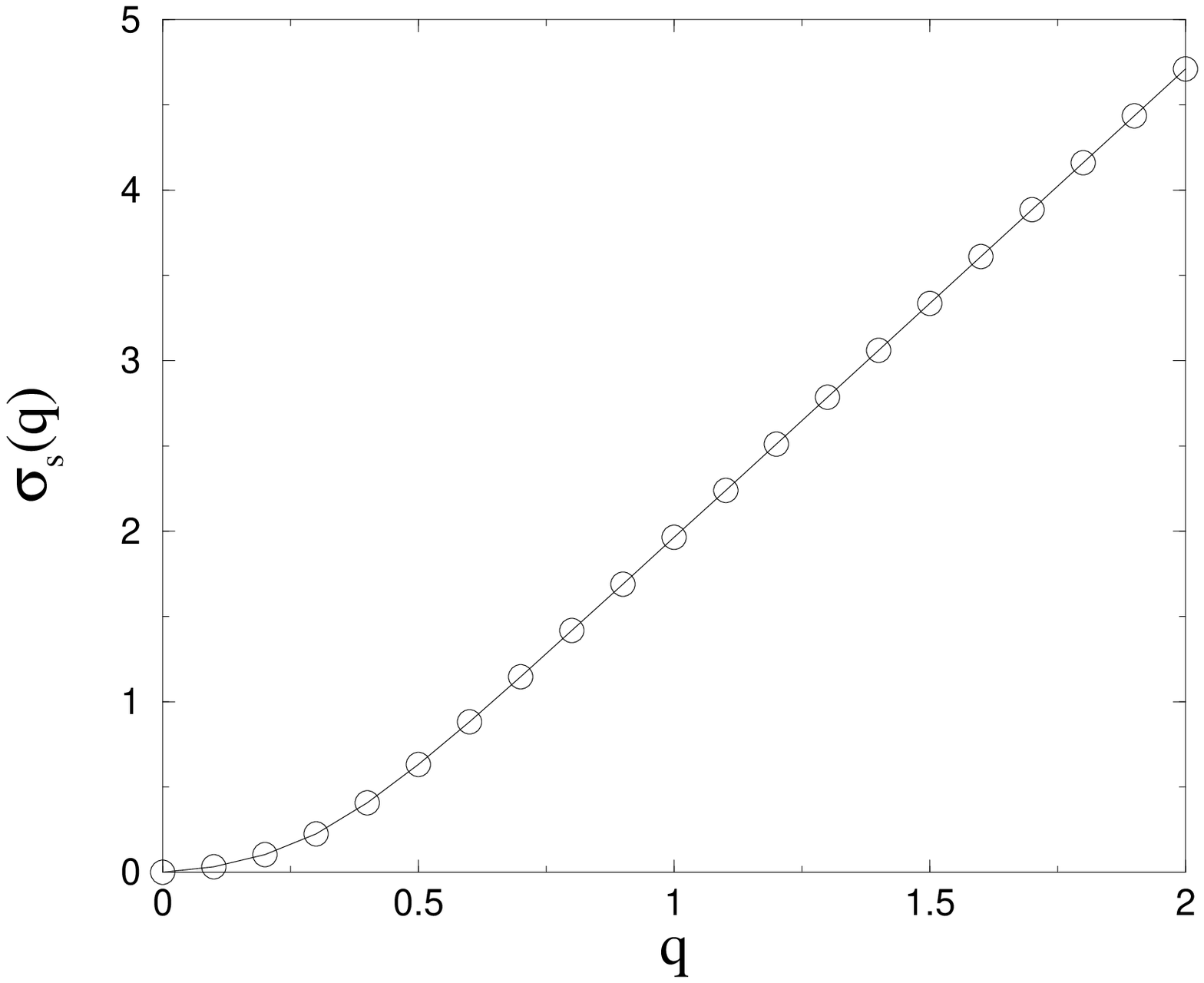, width=6cm}
    \hspace*{0.2cm}\epsfig{file=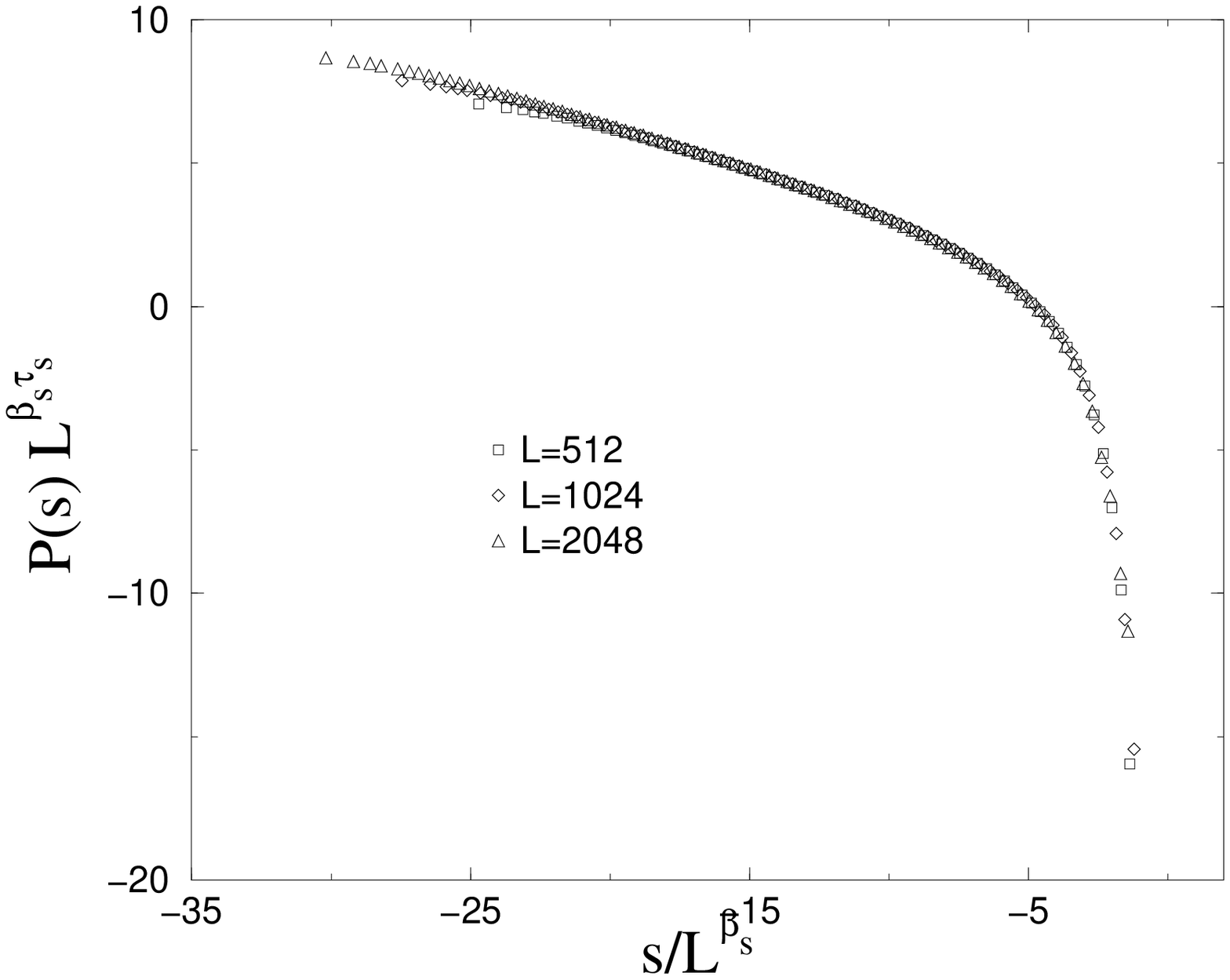, width=6cm}}
  \caption{(a) Plot of the moments spectrum for the distribution of 
toppling events $s$.The linear part has slope $2.73$.(b) Data collapse 
analysis for the avalanche size distribution. The values used for the 
critical exponents are $\tau_s=1.27$ and $\beta_s=2.7$.  }
  \label{sdist}
\end{figure}
The direct numerical determination of the exponents $\tau_x$ from the 
power law behavior of the probability distributions contains 
intrinsic bias due to the lower and upper cut-offs. This makes 
very difficult to get better than a $10\%$ accuracy. Extrapolations 
methods have been devised \cite{lubeck}, but the estimate of their 
accuracy is rather difficult. For these reasons, we use the more powerful 
moments analysis on the distribution $P(x,L)$, as also suggested recently 
by De Menech et al.\cite{att} on the BTW model.
We define the $q$-moment of $x$ on a lattice of size $L$ as
$\langle x^q\rangle_L=\int x^q P(x)dx$. 
If FSS hypothesis (Eq.~(\ref{pl})) is valid, at 
least in the asymptotic limit ($x\to\infty$), we can transform  
$z=x/L^{\beta_x}$ and obtain
\begin{equation}
\langle x^q\rangle_L= L^{\beta_x(q+1-\tau_x)}\int z^{q+\tau_x}{\cal G}(z)dz
\sim L^{\beta_x(q+1-\tau_x)},
\label{eq:xq}
\end{equation}
or in general $\langle x^q\rangle_L\sim L^{\sigma_x(q)}$.
The exponents $\sigma_x(q)$ can be obtained as the slope of the 
log-log plot of $<x^q>_L$ versus $L$. Using  Eq.~(\ref{eq:xq}), 
we obtain $\langle x^{q+1}\rangle_L/\langle x^q\rangle_L
\sim L^{\beta_x}$ or $\sigma_x(q+1)-\sigma_x(q)=\beta_x$,
so that the slope of $\sigma_x(q)$ as a function of $q$
is the cutoff exponent
$\beta_x =\partial\sigma_x(q)/\partial q$.
In Fig.s 1(a),2(a) and 3(a), we show the result obtained from the 
moments analysis of the distribution $P(s)$,$P(t)$ and $P(a)$, respectively.
In all cases, we get a clear linear behavior starting from $q\simeq 0.7$.
For smaller $q$ we observe deviations from standard FSS,
expected because
the integral in Eq.~(\ref{eq:xq}) is dominated by the lower cutoff 
for small $q$ moments. In particular, corrections to scaling  of the type 
$\langle x^q\rangle_L\sim L^{\sigma_x(q)}F(L)$
are important for $q \leq\tau_x-1$ and when  
$q\simeq \tau_x-1$, logarithmic corrections give rise 
to effective exponents up to very large  lattice sizes. 
\begin{figure}[t]
  \centerline{\epsfig{file=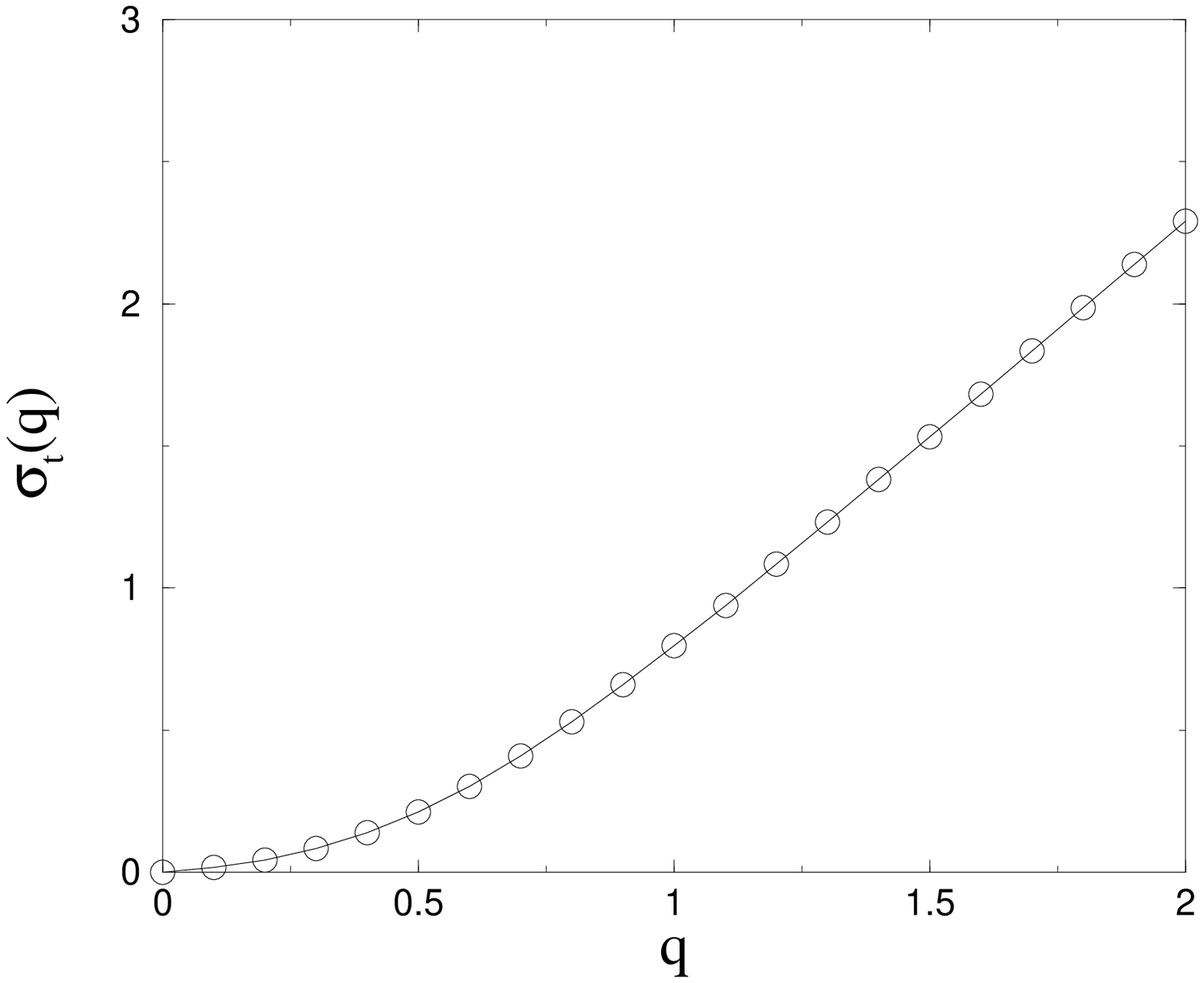, width=6cm}
    \hspace*{0.2cm}\epsfig{file=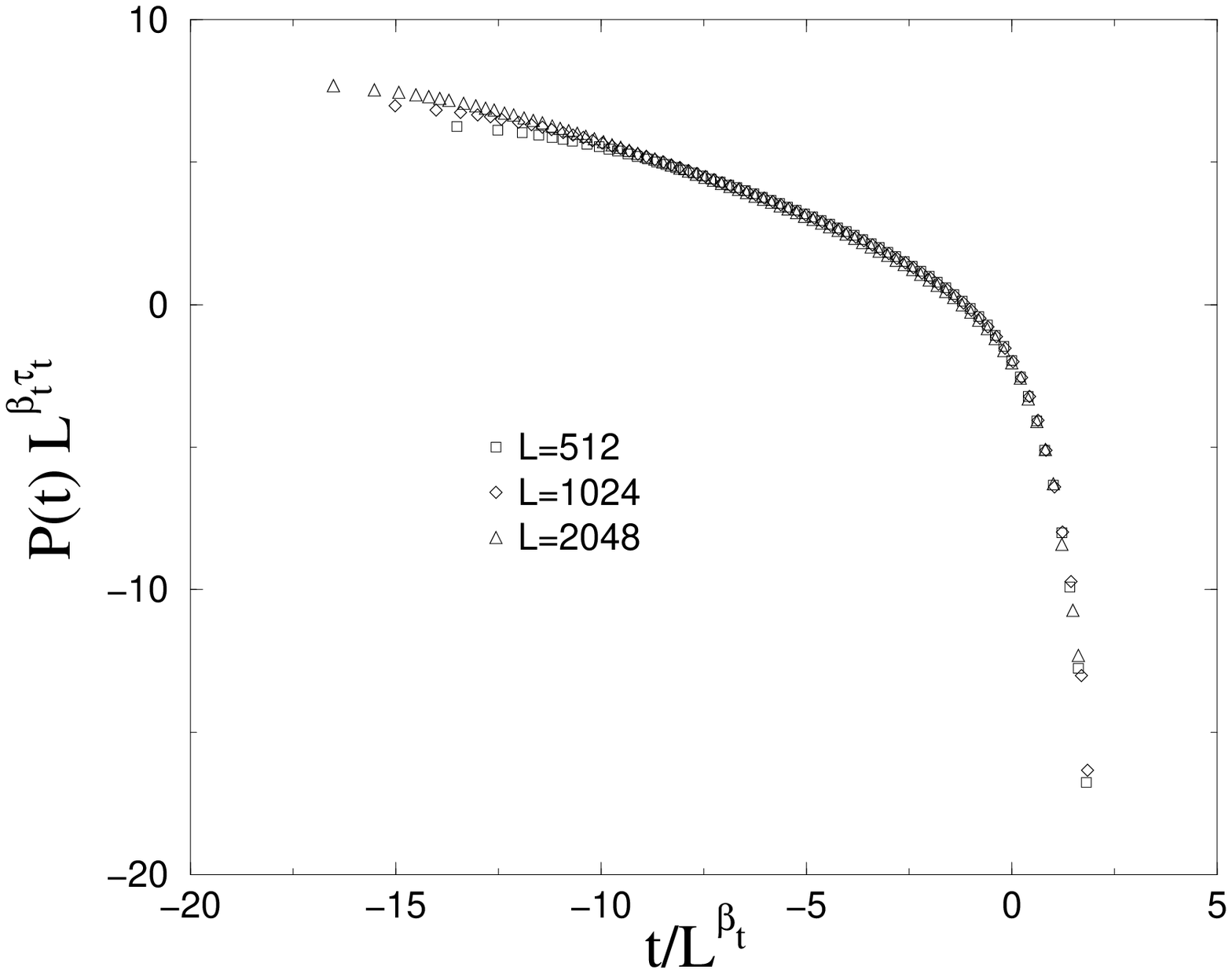, width=6cm}}
  \caption{(a) Plot of the moments spectrum for the distribution of 
avalanche time duration $t$.The linear part has slope $1.50$.(b) 
Data collapse analysis for the avalanche time duration distribution. 
The values used for the 
critical exponents are $\tau_t=1.5$ and $\beta_a=1.5$.  }
  \label{tdist}
\end{figure}
By measuring the slope of the linear part of  momentum spectra $\sigma_x(q)$,
we obtain the cut-offs exponents $\beta_s=2.73\pm0.02$,
$\beta_t=1.50\pm0.02$ and $\beta_a=2.02\pm0.02$. These exponents\cite{nota1}
are in good agreement with previous estimates for the manna model \cite{manna}.
If FSS is verified, we can compute the exponent $\tau_x$ from the scaling 
relation $(2-\tau_x)\beta_x=\sigma_x(1)$, that should be satisfied for 
enough large sizes. Using the values of $\beta_x$ reported in Table I and 
the values obtained for $\sigma_x(1)$ we  
find the exponents $\tau_s=1.27\pm0.01$, $\tau_t=1.50\pm0.01$ and  
$\tau_a=1.35\pm0.01$. Also in this case the values are in agreement 
with previous extrapolations or direct measurements\cite{manna,lubeck}.
\begin{figure}[t]
  \centerline{\epsfig{file=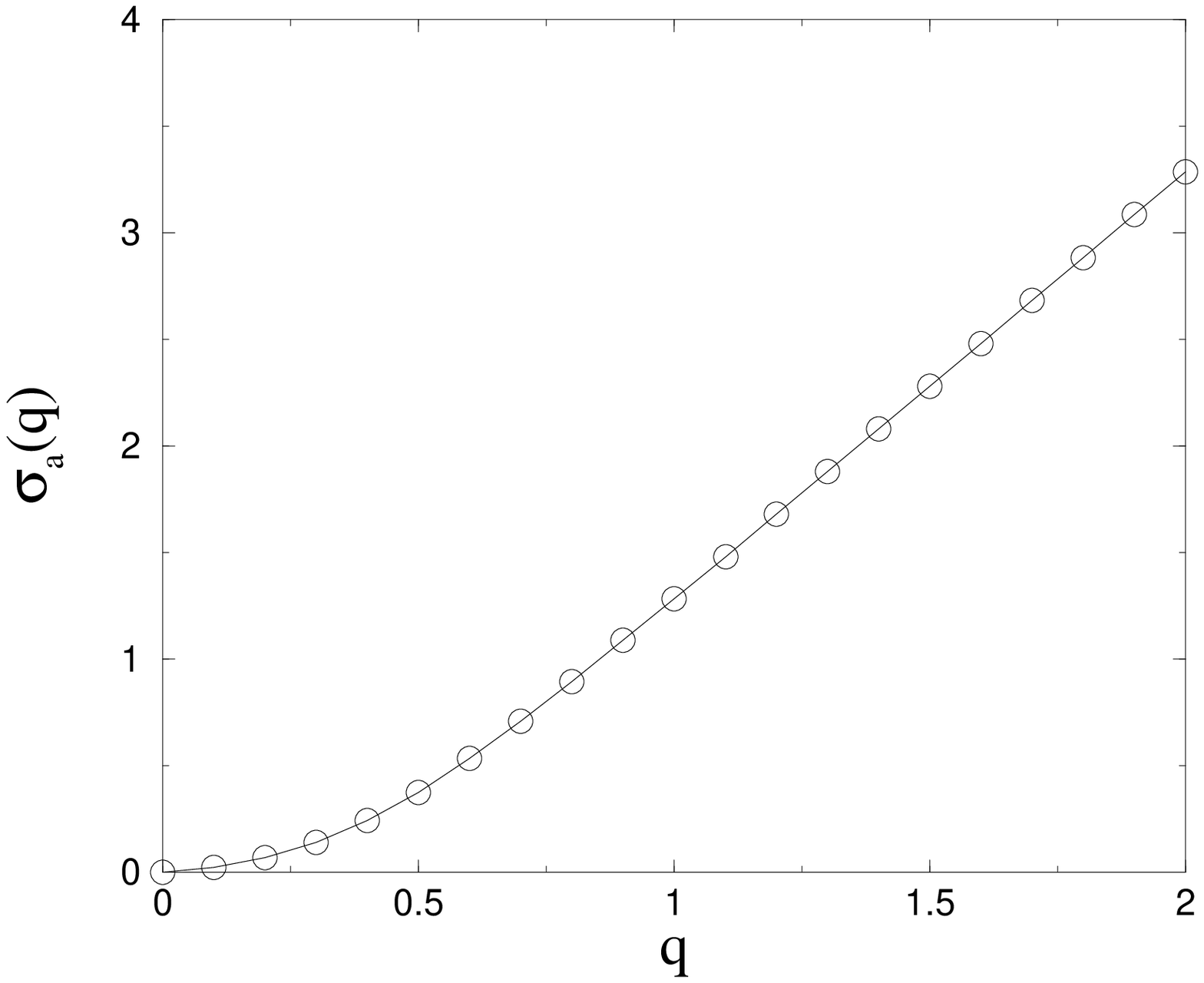, width=6cm}
    \hspace*{0.2cm}\epsfig{file=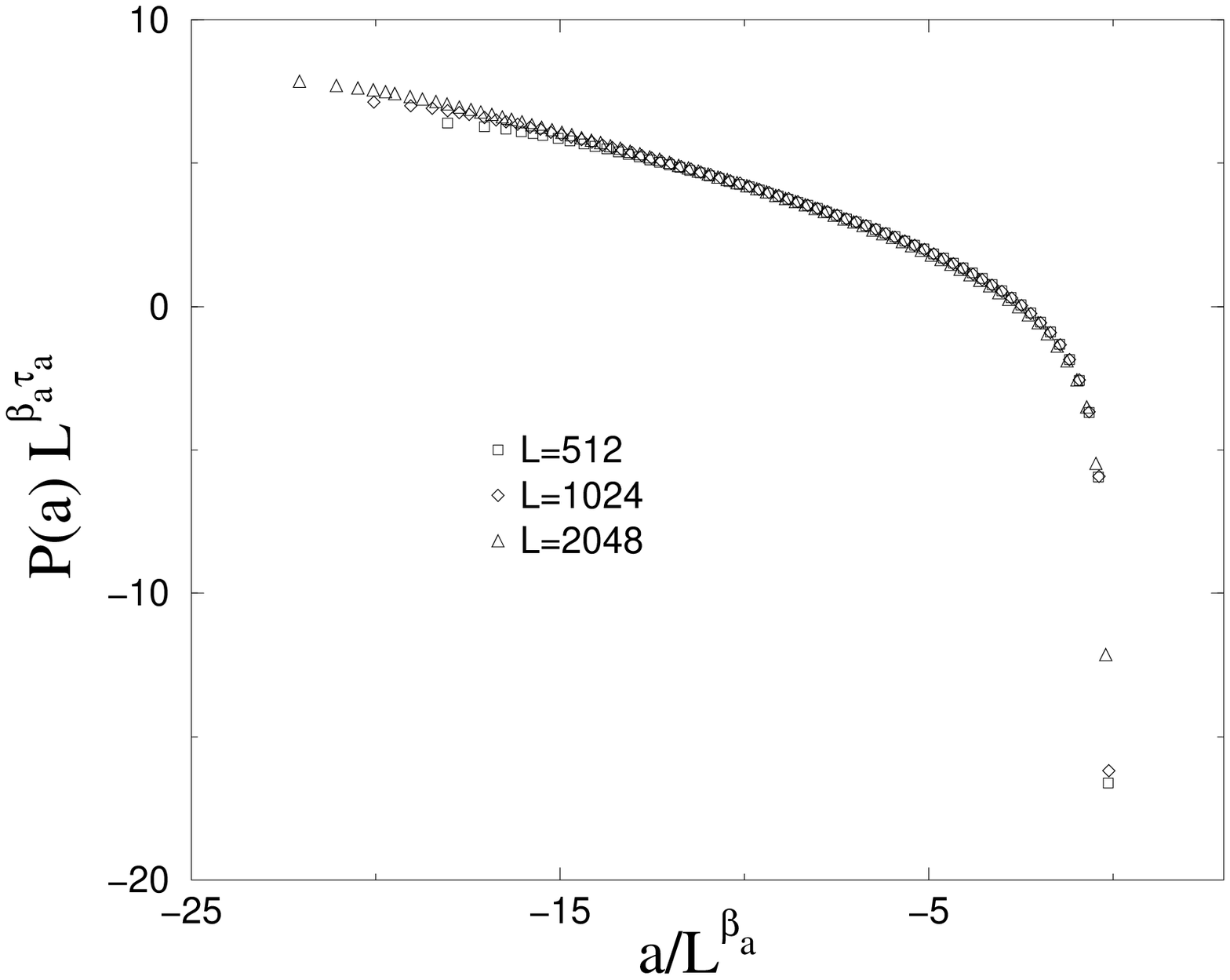, width=6cm}}
  \caption{(a) Plot of the moments spectrum for the distribution of 
avalanche area $a$.The linear part has slope $2.02$.(b) 
Data collapse analysis for the avalanche area distribution. 
The values used for the 
critical exponents are $\tau_a=1.35$ and $\beta_a=2.0$.  }
  \label{adist}
\end{figure}
As a final consistency test for the FSS framework we have to verify that 
we get data collapse for the distributions $P(x.L)$ by using the exponents 
obtained from the moments analysis. In fact, the FSS scenario states 
that rescaling  $q_x\equiv x/L^{\beta_x}$ 
and $P_{q_x}\equiv P(x,L)L^{\beta_x\tau_x}$, the data for different $L$
must collapse onto universal curves. In Fig.s~1(b), 2(b) and 3(b), we
show that very good  data collapses are obtained for all distributions.

In conclusion, we have reported extensive numerical simulations of
the Manna sandpile model in two dimensions \cite{manna}. 
We show that contrary to other sandpile
models, such as the BTW model \cite{btw}, where it is difficult
to obtain unambiguous results \cite{att,lubeck}, the FSS assumption 
is satisfied in the Manna model. In this way, 
we obtain the complete set of avalanche
critical exponents. Work is in progress to obtain comparable
results in $d=3,4$.


\begin{thebibliography}{10}
\bibitem{nota1}
        Note that the exponents $\beta_s$, $\beta_t$ 
        and $\beta_a$ are usually 
        reported in the literature as $D$, $z$ 
        and $d_f$, respectively.
\end{thebibliography}
\end{document}